\documentstyle[amstex,amssymb,righttag,verbatim,12pt]{article}
\newtheorem{th}{Theorem}
\newtheorem{pro}{Proposition}
\newtheorem{lem}{Lemma}
\newcommand{\sdet}{\operatorname{sdet}}
\begin{document}

\title{\sc Crum Transformation and
\\ Wronskian Type Solutions for\\
 Supersymmetric KdV Equation}

\author{Q. P. Liu\thanks{On leave of absence from
Beijing Graduate School, CUMT, Beijing 100083, China}
\thanks{Supported by {\em Beca para estancias temporales
de doctores y tecn\'ologos extranjeros en
Espa\~na: SB95-A01722297}}
   $\,$ and M. Ma\~nas\thanks{Partially supported by CICYT:
 proyecto PB92-019}\\
 Departamento de F\'\i sica Te\'orica, Universidad Complutense\\ 
E28040-Madrid, Spain.}

\date{}
\maketitle
\begin{abstract}
Darboux transformation is reconsidered
for the supersymmetric KdV system. 
By iterating the Darboux transformation, a supersymmetric extension
of  the Crum transformation is obtained
for the Manin-Radul SKdV equation, in doing so one gets
Wronskian superdeterminant representations for the
solutions. Particular examples provide us 
explicit supersymmetric extensions, super solitons,
of the standard soliton of the KdV equation. The KdV soliton
appears  as the body of the super soliton.
\end{abstract}
\newpage

\section{Introduction}

Supersymmetric integrable systems have attracted much attention
during last decade, and as a consequence a number of results
has been  established in this field. Thus, several well known
integrable systems, such as Kadomtsev-Petviashvili (KP) 
\cite{mr}, Korteweg-de Vries (KdV) \cite{mr}, Sine-Gordon \cite{ck}, Nonlinear
Schr\"odinger \cite{rk}, Ablowitz-Kaup-Newell-Segur/Zakharov-Shabat
\cite{mp} and Harry Dym \cite{liu1}
systems have been embedded into their supersymmetric counterparts.
 We notice that developing
this theory is not only interesting from a mathematical viewpoint,
but also may have physical relevance as toy models for
more realistic supersymmetric systems and in two dimensional
supersymmetric quantum gravity \cite{ag}.

Once these supersymmetric integrable models have been constructed one needs
to extend the non supersymmetric tools to construct solutions
to this supersymmetric framework. We are
particularly interested in Darboux transformation \cite{ms,o},
 which 
is very powerful to this end. Indeed, the initial
steps along this research line have been taken in \cite{liu} and 
there Darboux and binary Darboux transformations have been constructed
 for Manin-Radul super KdV (MRSKdV) and 
Manin-Radul-Mathieu super KdV systems. Furthermore, in \cite{lm}
vectorial Darboux transformations have been given for the
MRSKdV and explicit soliton type solutions were presented. 

In this  paper we give 
a Crum type transformation \cite{c,wks}
for  the MRSKdV system and its Lax operator, 
a supersymmetric extension
of the stationary Schr\"odinger operator. In doing so,  given
any solution of the system and wave functions solving the
associated linear system,  we obtain
new solutions represented in terms of
superdeterminants of matrices of Wronski type.

The paper is organized as follows. In \S 2
we consider the iteration of the Darboux transformation
of \cite{liu} obtaining a Crum type transformation
and associated Wronskian superdeterminant expressions
for the new solutions, we also see how this transformation reduces down
to the classical case of KdV equation. 
In \S 3 we present some simple examples,
constructing a family of solutions that can be considered as 
a super soliton
solution because it is a supersymmetrized version of the standard
soliton. Indeed, the KdV soliton solution appears as the body of 
the super soliton, which in turn has its soul exponentially localized.

\section{Darboux Transformation for 
MRSKdV: Iteration and Reduction}
The MRSKdV system was introduced in the 
context of supersymmetric KP hierarchy \cite{mr} and it reads as
\begin{equation}\label{mr}
\begin{aligned}
\partial_t \alpha&={1\over 4}\partial (\partial^2\alpha+3\alpha D\alpha
+6\alpha u),\\
 \partial_t u&={1\over 4}\partial (\partial^2u+3u^2+3\alpha Du),
\end{aligned}
\end{equation}
where we use the notation 
$\partial f:=\partial f/\partial x$, $\partial_t f:=\partial f/\partial t$, 
$x, t\in \Bbb C$ and
$ D$ is the super derivation defined by $D:=\partial_{\vartheta}+
\vartheta \partial$ with
$\vartheta$ a Grassmann odd variable.

It is known that the  following linear system 
\begin{equation}\label{linear}
\begin{aligned} 
(L-\lambda)\psi&:=(\partial^2+\alpha D +u-\lambda) \psi=0,\\
(\partial_t-M)\psi&:=\big[\partial_t-{1\over 2}\alpha\partial D-\lambda\partial-{1\over2}u\partial
+{1\over 4}(\partial\alpha) D+{1\over 4}(\partial u)\big]\psi=0,
\end{aligned}
\end{equation}
has as its compatibility condition
\[
[\partial_t-M,L]=0,
\]
the system (\ref{mr}). Thus,  $L$ is the Lax operator
for the MRSKdV system, and $L\psi=\lambda\psi$ is a supersymmetric extension
of the stationary Schr\"odinger equation.

The Darboux transformation for the MRSKdV equation \cite{liu},
 that we will iterate in this paper, is
\begin{pro} Let $\psi$ be a solution of
\begin{equation}\label{l1}
\begin{aligned}
L\psi&=\lambda\psi,\\
\partial_t\psi&=M\psi,
\end{aligned}
\end{equation}
and $\theta_0$ be a particular solution with  $\lambda=\lambda_0$.
Then, the quantities defined by
\begin{align*}
{\hat{\psi}}: &=(D+\delta_0)\psi,
 \qquad \delta_0:=-{D\theta_0\over \theta_0}, \quad ({\theta_0}: \text{even})\\
\hat{\alpha}:&= -\alpha -2\partial\delta_0,\\
\hat{u}: &=u+(D\alpha)+2\delta_0(\alpha+\partial\delta_0)
\end{align*}
satisfy
\begin{align*}
\hat L\hat\psi&=\lambda\hat\psi,\\
\partial_t\hat \psi&=\hat M\hat\psi,
\end{align*}
where $\hat L$ and $\hat M$ are obtained from $L$ and $M$ by replacing
$\alpha$ and $u$ with $\hat\alpha$ and $\hat u$, respectively.
\end{pro}

As a consequence of this Proposition we conclude that $\hat u$ and $\hat\alpha$
are new solutions of the MRSKdV system (\ref{mr}).
We remark that, as usual, the Darboux transformation 
can be viewed as a gauge transformation:
\begin{align*}
\psi&\to T_0\psi,\\
L&\to \hat L=T_0LT_0^{-1},\\
M&\to \hat M=\partial_tT_0\cdot T_0^{-1}+T_0MT_0^{-1},\\
T_0&:=D+\delta_0.
\end{align*}

To obtain Crum type transformation, 
let us start with $n$  solutions $\theta_i$, 
$i =0,...,n-1$, of  equation (\ref{linear})
with eigenvalues as $\lambda=k_i$, $i=0,..., n-1$.
 To make sense, we choose
the $\theta_i $ in such way that its index indicates its
parity: those with even indices are even and with odd indices
are odd variables. We use $\theta_0$ to do our first
step  transformation and then $\theta_i$, $ i=1,\dots , n-1$, are transformed to new solutions $\theta_i[1]$
of the transformed linear equation and $\theta_0$
goes to zero. Next step can be effected by
using $\theta_1[1]$ to form a Darboux operator and at this
time $\theta_1[1]$ is lost. We can continue this
iteration process until all the seeds are mapped to zero. In
this way, we have
\begin{pro}
Let $\theta_i$, $i=0,\dots , n-1$, be solutions of
the linear system (\ref{linear}) with 
$\lambda=k_i$, $i=0,\dots ,n-1$, and parities $ p(\theta_i)=(-1)^i$,
 then after $n$ iterations of the
Darboux transformation of Proposition 1, one obtains a new Lax
operator
\[
{\hat{L}}=T_nLT_{n}^{-1}, \quad T_n=D^n +\sum_{i=0}^{n-1}a_iD^i,
\]
where the coefficients $a_i$ of the gauge operator $T_n$ are defined
by
\begin{equation}\label{l2}
(D^n +\sum_{i=0}^{n-1}a_iD^i)\theta_j=0, \qquad j=0,\dots , n-1.
\end{equation}
\end{pro}

{\bf Proof}: With 
$\theta_i$, $i=0,\dots ,n-1$, we may perform  $n$ Darboux
 transformations of Proposition 1 step by step. 
Indeed, in the first step we transform using $\theta_0$, obtaining:
\begin{align*}
&\psi[1]:=(D+\delta_0)\psi, \qquad
 \delta_0:=-{D\theta_0\over\theta_0}.\\
&L[1]:=(D+\delta_0)L(D+\delta_0)^{-1},
\end{align*}
and 
\begin{align*}
&\theta_i[1]:=(D+\delta_0)\theta_i, \qquad i=1,\dots, n-1,
\end{align*}
notice that $\theta_0[1]=(D+\delta_0)\theta_0=0$.

Using $\theta_1[1]$ to do our next step transformation we get
\begin{align*}
&\psi[2]:=(D+\delta_1)\psi[1], \qquad \delta_1:=-(\theta_1[1])^{-1}
D\theta_1[1],\\
&L[2]:=(D+\delta_1)L[1](D+\delta_1)^{-1},\\
&\theta_i[2]:=(D+\delta_1)\theta_i[1],\qquad i=2,\dots , n-1,
\end{align*}
and at this time we have $\theta_1[2]=0$.

Combining these two steps, we get
\begin{align*}
\psi[2]&=(D+\delta_1)(D+\delta_0)\psi,\\
L[2]&=(D+\delta_1)(D+\delta_0)L(D+\delta_0)^{-1}(D+\delta_1)^{-1}.
\end{align*}

Following this way, we iterate the Darboux transformations until all
the seeds, $\{\theta_i\}_{i=0}^{n-1}$, are used up. 
It is clear that our composed gauge operator has the following form
\begin{align*}
&T_n=(D+\delta_{n-1})\cdots(D+\delta_0)=D^n+a_{n-1}D^{n-1}+\cdots +
a_0,
\end{align*}
with the property $T_n(\theta_j)=0$. This in turn
determines the coefficients $a_i$, $i=0,\dots,n-1$. $\Box$

The explicit form of the transformed  
field variables is given by ${\hat{L}}=T_nLT_{n}^{-1}$:
\begin{lem}
The new fields $\hat\alpha$ and $\hat u$ can be written as
\begin{equation}\label{te}
\begin{aligned}
\hat{\alpha}&=(-1)^n\alpha-2\partial a_{n-1},\\
{\hat{u}}&=u-2\partial a_{n-2}-a_{n-1}((-1)^n \alpha +{\hat{\alpha}})+
\frac{1-(-1)^n}{2}D\alpha.
\end{aligned}
\end{equation}
\end{lem}

Next, we  solve the linear  system (\ref{l2}) and 
get the explicit solutions in terms of superdeterminants.
Since the cases for $ n $ even or odd are rather different,
 we consider them separately.

For $n=2k$, we denote by
\begin{align*}
{\boldsymbol a}^{(0)}:&=(a_0, a_2,\dots, a_{2k-2}),\qquad
{\boldsymbol a}^{(1)}:=(a_1, a_3,\dots, a_{2k-1}),\\
{\boldsymbol\theta}^{(0)}:&=(\theta_0,\theta_2,\dots,\theta_{2k-2}), \qquad
{\boldsymbol\theta}^{(1)}:=(\theta_1,\theta_3,\dots,\theta_{2k-1}), \\
{\boldsymbol b}^{(i)}:&=\partial^k{\boldsymbol\theta}^{(i)},\qquad
W^{(i)}:=\left(\begin{array}{c}{\boldsymbol\theta}^{(i)}\\
\partial{\boldsymbol\theta}^{(i)}\\ \vdots\\ \partial^{k-1}{\boldsymbol\theta}^{(i)}
\end{array}\right),\qquad i=0,1.
\end{align*}
Then,  our linear  system (\ref{l2}) can be now formulated as
\begin{equation}\label{l3}
({\boldsymbol a}^{(0)}, {\boldsymbol a}^{(1)})\cal W
=-({\boldsymbol b}^{(0)}, {\boldsymbol b}^{(1)}),
\end{equation}
where
\[
\cal W:=\left(\begin{array}{cc}
W^{(0)}&W^{(1)}\\DW^{(0)}&DW^{(1)}\end{array}\right).
\]
It should be noticed that the supermatrix $\cal W$
is even and  has a Wronski type structure.

We recall some facts about supermatrices \cite{ber,dw}.
Given an even matrix, say $\cal M=\big(\begin{smallmatrix}
 A & B\\ C &D
\end{smallmatrix}\big)$, its inverse is
\[
\cal M^{-1}=\left(\begin{array}{cc}(A-BD^{-1}C)^{-1}&
-A^{-1}B(D-CA^{-1}B)^{-1}\\
-D^{-1}C(A-BD^{-1}C)^{-1}&
(D-CA^{-1}B)^{-1}\end{array}\right),
\]
and  its  Berezinian or superdeterminant is
\[
\sdet\cal M=\frac{\det \left(A -B D^{-1}C\right)}
{\det D}=\frac{\det A}{\det\left(D- C A^{-1} B\right)}.
\]

Now, we have
\begin{lem}
For $n=2k$, one can write
\[
a_{2k-2}=-\frac{\sdet\hat{\cal W}}{\sdet{\cal W}},\qquad
a_{2k-1}=D \ln\, \sdet {\cal W},
\]
here 
\[
\hat{\cal W}=\left(\begin{array}{cc} {\hat W}^{(0)}&{\hat W}^{(1)}\\
DW^{(0)}& DW^{(1)}\end{array}\right)
\]
where
$ {\hat{W}}^{(0)} $ and $ {\hat{W}}^{(1)}$ are obtained from the matrices
$W^{(0)}$ and $W^{(1)}$ by replacing the last rows with 
${\boldsymbol b}^{(0)}$ and ${\boldsymbol b}^{(1)}$, respectively.
\end{lem} 
{\bf Proof}: 
Multiplying (\ref{l3}) by $\cal W^{-1}$ one finds
\begin{align*}
{\boldsymbol a}^{(0)}&=-\left({\boldsymbol b}^{(0)}-
{\boldsymbol b}^{(1)}(DW^{(1)})^{-1}DW^{(0)}\right)
\left(W^{(0)}-W^{(1)}(DW^{(1)})^{-1}DW^{(0)}\right)^{-1},\\
{\boldsymbol a}^{(1)}&=-\left({\boldsymbol b}^{(1)}-
{\boldsymbol b}^{(0)}(W^{(0)})^{-1}W^{(1)}\right)
\left(DW^{(1)}-(DW^{(0)})(W^{(0)})^{-1}W^{(1)}\right)^{-1}.
\end{align*}
Noticing that both 
\[
\left(W^{(0)}-W^{(1)}(DW^{(1)})^{-1}DW^{(0)}\right)\text{ and }
\left(DW^{(1)}-(DW^{(0)})(W^{(0)})^{-1}W^{(1)}\right)
\]
 are even matrices, we  use the Cramer's rule to obtain
\[
a_{2k-2}=-\frac{\det \left({\hat{W}}^{(0)}-{\hat{W}}^{(1)}(DW^{(1)})^{-1}
DW^{(0)}\right)}{\det \left(W^{(0)}-W^{(1)}(DW^{(1)})^{-1}DW^{(0)}\right)}
=-\frac{\sdet \hat{\cal W}}{\sdet {\cal W}}.
\]

Similarly, for $a_{2k-1}$ we have the following expression 
\[
a_{2k-1}=-\frac{\det \left({D\bar{W}}^{(1)}-(D{\bar{W}}^{(0)})(W^{(0)})^{-1}
W^{(1)}\right)}{\det 
\left(DW^{(1)}-(DW^{(0)})(W^{(0)})^{-1}W^{(1)}\right)},
\]
where ${\bar W}^{(i)}$ is obtained from $W^{(i)}$ with its last row
replaced by its $D$ derivation, that is, 
$D^{2k-1}{\boldsymbol \theta}^{(i)}$.
We will  show next that one can write $ a_{2k-1}=D\ln \,\sdet {\cal W}$.

Introduce the notation 
\[
V:=DW^{(1)}-(DW^{(0)})(W^{(0)})^{-1}W^{(1)},
\]
and denote by ${\boldsymbol v}_j$ the $j$th row of the matrix $V$.
 Then, since
\[
(\sdet {\cal W})^{-1}=\frac{\det V}
{\det W^{(0)}},
\]
we have 
\begin{align*}
D\big((\sdet {\cal W})^{-1}\big)&=\frac{D\det V}{\det W^{(0)}}-\frac{D\ln\det W^{(0)}}{\sdet {\cal W}}\\
              &=\frac{D\det V}{\det W^{(0)}}-
\frac{\text{Tr}\big( (DW^{(0)})(W^{(0)})^{-1}\big)}{\sdet {\cal W}}.
\end{align*}
Notice that
\begin{align*}
D\det V&=\sum^{k}_{j=1}\det\left(V-\boldsymbol e_j\otimes(\boldsymbol v_j-
D\boldsymbol v_j)\right),\\
DV&=\partial W^{(1)}-(\partial W^{(0)})(W^{(0)})^{-1}W^{(1)}+
(DW^{(0)}) (W^{(0)})^{-1}V ,\\
\partial W^{(j)}&=\Lambda W^{(j)}+\boldsymbol e_k\otimes\boldsymbol b^{(j)}
\end{align*}
where $\boldsymbol e_i=(\delta_{ij})$ is a column vector with all its
entries vanishing except for the $i$th one,
and $\Lambda=(\delta_{i,i+1})$ is the shift matrix.

Hence, we obtain

\[
D{\boldsymbol v}_j=\delta_{j,k} \left({\boldsymbol b}^{(1)}-{\boldsymbol b}^{(0)} (W^{(0)})^{-1}W^{(1)}\right)
+{\boldsymbol c} _j V, \qquad j=1,\dots, k
\]
where we denote by ${\boldsymbol c}_j$  the $j$th row of the
matrix $(DW^{(0)})(W^{(0)})^{-1}$.

With the help of above formulae  we have
\begin{align*}
D\det V=&\det\big(V-\boldsymbol e_k\otimes\big[\boldsymbol v_k-
(\boldsymbol b^{(1)}-
\boldsymbol b^{(0)}(W^{(0)})^{-1}W^{(1)})\big]\big)\\ &
+\sum_{j=1}^k\det(1-\boldsymbol e_j\otimes(\boldsymbol e_j-
\boldsymbol c_j))\det V
\\
=&\det\left(D{\bar W}^{(1)}-
(D{\bar W}^{(0)})(W^{(0)})^{-1}W^{(1)}\right)+
\text{Tr}\left((DW^{(0)})(W^{(0)})^{-1}\right)\det V,
\end{align*}
so that 
\[
D\big((\sdet {\cal W})^{-1}\big)=\frac{\det\left(D{\bar W}^{(1)}-
(D{\bar W}^{(0)})(W^{(0)})^{-1}W^{(1)}\right)}{\det W^{(0)}}
\]
which leads to our claimed formula for $a_{2k-1}$.
$\Box$

For the $n=2k+1$ case, we need to introduce a different set
of notations:
\begin{align*}
{\boldsymbol a}^{(1)}&=(a_0, a_2,\dots, a_{2k}),\qquad
{\boldsymbol a}^{(0)}=(a_1, a_3,\dots, a_{2k-1}),\\
{\boldsymbol \theta}^{(0)}&=(\theta_0,\theta_2, \dots, \theta_{2k}),\qquad
{\boldsymbol \theta}^{(1)}=(\theta_1,\theta_3, \dots, \theta_{2k-1}),\\
{\boldsymbol b}^{(i)}&=D^{2k+1}{\boldsymbol \theta}^{(j)},
\quad i\neq j,\quad
W^{(i)}=\left(\begin{array}{c} {\boldsymbol\theta}^{(i)}\\
\partial{\boldsymbol\theta}^{(i)}\\ \vdots \\ \partial^k {\boldsymbol\theta}^{(i)}
\end{array}\right), \quad i,j=0, 1,
\end{align*}
then our linear  system (\ref{l2}) reads as
\[
({\boldsymbol a}^{(1)}, {\boldsymbol a}^{(0)}) \left(\begin{array}{cc}
W^{(0)}&W^{(1)}\\ D {\tilde W}^{(0)}&D{\tilde W}^{(1)}\end{array}\right)
=-({\boldsymbol b}^{(1)}, {\boldsymbol b}^{(0)}),
\]
where ${\tilde W}^{(i)}$ is the $W^{(i)}$ with the last row removed.

We introduce two block matrices
\[
\cal W=\left(\begin{array}{cc}
W^{(0)}&W^{(1)}\\ D {\tilde W}^{(0)}&D{\tilde W}^{(1)}\end{array}\right),\qquad
\check{\cal W}=\left(\begin{array}{cc}
W^{(0)}&W^{(1)}\\ D {\check W}^{(0)}&D{\check W}^{(1)}\end{array}\right),
\]
where $\check W^{(i)}$ is the matrix
$W^{(i)}$ with its second to last row removed.

\begin{lem}
For $n=2k+1$, one has
\begin{align*}
a_{2k-1}&=-\frac{\sdet {\cal W }}{\sdet \check {\cal W}},\\
a_{2k}&=-\frac{\det ({\hat W^{(0)}}-{\hat W}^{(1)}(D{\tilde W}^{(1)})^{-1}
(D{\tilde W}^{(0)}))}{\det( W^{(0)}-W^{(1)}(D{\tilde W}^{(1)})^{-1}
D{\tilde W}^{(0)})},
\end{align*}
where ${\hat{W}}^{(i)}$ is the matrix
$W^{(i)}$ with its last row replaced by the vector
${\boldsymbol b}^{(j)}$, $i,j=0, 1,$, $i\neq j$ .
\end{lem}

{\bf Proof:} This lemma follows easily form Cramer's rule.$\Box$

\newpage

Now, from the last three lemmas one can easily arrive to the main 
result of this paper:
\begin{th}
Let $\alpha,u$ be a seed solution of (\ref{mr}) and
$\{\theta_j\}_{j=0}^{n-1}$ be a set of $n$ solutions of
the associated linear system (\ref{linear}), such that
the parity is $p(\theta_j)=(-1)^j$.
Then:
\begin{itemize}
\item[\bf (i)]
If $n=2k$,  we have  new solutions $\hat\alpha$, $\hat u$ of
(\ref{mr}) given by
\begin{align*}
\hat\alpha&=\alpha-2 D^3\ln\,\sdet{\cal W},\\
\hat u&=u+2\partial\Big(\frac{\sdet\hat{\cal W}}{\sdet {\cal W}}\Big)+
(\alpha+\hat\alpha) D\ln\,\sdet{\cal W}.
\end{align*}
\item[\bf (ii)]
If $n=2k+1$,  we have new solutions $\hat\alpha$, $\hat u$ of
(\ref{mr}) given by
\begin{align*}
\hat\alpha=&\alpha+2\partial \Biggl(\frac{\det ({\hat W^{(0)}}-{\hat W}^{(1)}(D{\tilde W}^{(1)})^{-1}
D{\tilde W}^{(0)})}{\det( W^{(0)}-W^{(1)}(D{\tilde W}^{(1)})^{-1}
D{\tilde W}^{(0)})}   \Biggr),\\
\hat u=&u+2\partial\Big(\frac{\sdet{\cal W} }{\sdet \check {\cal W}}\Big)
+D\alpha\\ &-
(\alpha-\hat\alpha)\Big(\frac{\det ({\hat W^{(0)}}-{\hat W}^{(1)}(D{\tilde W}^{(1)})^{-1}
(D{\tilde W}^{(0)}))}{\det( W^{(0)}-W^{(1)}(D{\tilde W}^{(1)})^{-1}
D{\tilde W}^{(0)})}   \Big).
\end{align*}
\end{itemize}
\end{th}

Some remarks are in order here
\begin{itemize}
\item[\bf (i)] The proof of Lemma {\bf 2}  given above 
is inspired by the one in \cite{uy}.
\item[\bf (ii)] In the $n=2k+1$ case,
 it is not possible to 
write $a_{2k}$ in terms of superdeterminants, this is so because
 the block structure
of the matrix is not preserved after using Cramer's rule.
\item[\bf (iii)] 
The elegant representation for $a_{2k-1}$ obtained in the even case is
lost in the odd case.
\end{itemize}

\paragraph{Reduction to KdV}
Eq. (\ref{mr}) reduces to the KdV equation when $\alpha=0$.
Hence, it is natural to ask whether our iterated Darboux transformation
reduces down to a
 Darboux transformation of the KdV equation. Actually, that aim is
achieved when we take, $n=2k$,  $a_{2i-1}=0$ with $\alpha=0$.
 Under this condition the linear system (\ref{l3}) breaks into:
\begin{equation}
{\boldsymbol a}^{(0)}W^{(0)}=-{\boldsymbol b}^{(0)},\qquad
{\boldsymbol a}^{(0)}W^{(1)}=-{\boldsymbol b}^{(1)}.
\end{equation}
Thus, while the first equation leads to the well known formula for the
KdV system \cite{ms}, the second one is a constraint for $\theta_i$, $i=0,\dots,n-1$. 
This constraint can be easily solved with the choice
\[
\theta_{2i}(x,t,\vartheta)=\theta_{2i}(x,t), \qquad
\theta_{2i-1}(x,t,\vartheta)=\vartheta\theta_{2i}(x,t)
\]
so that $W^{(1)}=\vartheta W^{(0)}$ and ${\boldsymbol b}^{(1)}=\vartheta {\boldsymbol b}^{(0)}$,
therefore the equation ${\boldsymbol a}^{(0)}W^{(1)}=-{\boldsymbol b}^{(1)}$ holds
whenever ${\boldsymbol a}^{(0)}W^{(0)}=-{\boldsymbol b}^{(0)}$ does.

\section{Super Solitons}

In this section we present some explicit examples obtained by dressing
the zero background $u=\alpha=0$ for $n=2$.
The general
solutions of the linear system (\ref{l2}) are:
\[
\theta_0=c_{+}\exp(\eta)+c_{-}\exp(-\eta),\qquad
\theta_1=\gamma_{+}\exp(\eta)+\gamma_{-}\exp(-\eta),
\]
where $\eta=k x+k^{3}t$, $k\in\Bbb C$ and
 $c_{\pm}$ are even and $\gamma_{\pm}$ are odd
and
\[
c_{\pm}(\vartheta)=c_{\pm}^{(0)}+\vartheta c_{\pm}^{(1)},\qquad
\gamma_{\pm}(\vartheta)=\gamma_{\pm}^{(1)}+\vartheta \gamma_{\pm}^{(0)},
\]
here the superfix indicates the parity.

In order to simplify the final expressions we take $c_\pm^{(0)},
\gamma_\pm^{(0)}\in\Bbb C$.
One can show that 
\begin{align*}
\sdet \cal W=&
\frac{c_{+}^{(0)}\exp(\eta)+c_{-}^{(0)}\exp(-\eta)}
{\gamma_{+}^{(0)}\exp(\eta)+\gamma_{-}^{(0)}\exp(-\eta)}
\\
&+\frac{\left(c_{+}^{(1)}\exp(\eta)+c_{-}^{(1)}\exp(-\eta)\right)
\left(\gamma_{+}^{(1)}\exp(\eta)+\gamma_{-}^{(1)}\exp(-\eta)\right)}
{\left(\gamma_{+}^{(0)}\exp(\eta)+\gamma_{-}^{(0)}\exp(-\eta)\right)^2}
\\
&+\frac{2\vartheta k\left(c^{(0)}_{+}\gamma_{-}^{(1)}-c_{-}^{(0)}\gamma_{+}^{(1)}\right)}
{\left(\gamma_{+}^{(0)}\exp(\eta)+\gamma_{-}^{(0)}\exp(-\eta)\right)^2}
-\frac{4\vartheta k  \gamma_{-}^{(1)}\gamma_{+}^{(1)}
\left( c_{+}^{(1)}\exp(\eta)+c_{-}^{(1)}\exp(-\eta)\right)}
{\left(\gamma_{+}^{(0)}\exp(\eta)+\gamma_{-}^{(0)}\exp(-\eta)\right)^3}.
\end{align*}
Hence, we have
\begin{align*}
a_0&=f-k\big(\gamma^{(1)}_-\gamma^{(1)}_+
+\vartheta(\gamma^{(1)}_+\gamma^{(0)}_--\gamma^{(1)}_-
\gamma^{(0)}_+)\big)g-
\vartheta k\big(c^{(0)}_+\gamma^{(1)}_--c^{(0)}_-\gamma^{(1)}_+\big)fg,\\
a_1&=\big(k(c^{(0)}_+\gamma^{(1)}_--c^{(0)}_-\gamma^{(1)}_+)+
\vartheta(c^{(0)}_+\gamma^{(0)}_--c^{(0)}_-
\gamma^{(0)}_+)\big)g,
\end{align*}
where
\begin{align*}
f&:=-k\left(
\frac{c_{+}^{(0)}\exp(\eta)-c_{-}^{(0)}\exp(-\eta)}
{c_{+}^{(0)}\exp(\eta)+c_{-}^{(0)}\exp(-\eta)}
\right),\\
g&:=\frac{2}{\left(c_{+}^{(0)}\exp(\eta)+c_{-}^{(0)}\exp(-\eta)\right)
\left(\gamma_{+}^{(0)}\exp(\eta)+\gamma_{-}^{(0)}\exp(-\eta)\right)}.
\end{align*}
Our solution is
\[
{\hat \alpha}=-2\partial a_{1}, \qquad 
{\hat u}=-2\partial a_{0}.
\]
Notice that our solution can be understood as a super 
soliton which has the KdV soliton, $-2\partial f$, as its body, and that
the choice $c_{+}^{(0)}=\gamma_{+}^{(0)}$ and 
$c_{-}^{(0)}=\gamma_{-}^{(0)}$ gives the solution found in \cite{imm}.

{\bf Acknowledgement} 
We should like to thank Allan Fordy
and Kimio Ueno for making the reference \cite{uy} available to us.


\begin{thebibliography}{99}


\bibitem{ag} L. Alvarez-Gaum\'e, H. Itoyama, J. L. Ma\~nes
 and A. Zadra, {\em Int. J. Mod. Phys.} {\bf A7} (1992) 5337;
 L. Alvarez-Gaum\'e, K. Becker, M. Berker , R. Emparan and
J. L. Ma\~nes, {\em Int. J. Mod. Phys.} {\bf A8} (1993) 2297. 

\bibitem{ber} F. A. Berezin, {\em Introduction to Superanalysis}, (D. Reidel Publishing
      Company, 1987).
\bibitem{ck} M. Chaichian and P. P. Kulish, {\em Phys. Lett.}
      {\bf B77} (1978) 413.

\bibitem{c} M. Crum, { \em Q. J. Math.} {\bf 6} (1955) 121.

\bibitem{dw} B. DeWitt, {\em Supermanifolds}, (Cambridge
University Press, 1984).

\bibitem{imm} L. A. Ibort, L. Mart\'\i nez Alonso and E. Medina,
     {\em J. Math. Phys.} (in press).


\bibitem{liu1} Q. P. Liu, {J. Phys. A: Math. Gen.} {\bf 28} (1995) L245.

\bibitem{liu} Q. P. Liu, {\em Lett. Math. Phys.} {\bf 35} (1995) 115.

\bibitem{lm} Q. P. Liu and M. Ma\~nas, Darboux Transformations
     for the Manin-Radul Supersymmetric KdV Equation, preprint(1996).

\bibitem{mr} Yu. I. Manin and A. O. Radul, {\em Commun. Math. Phys.}
     {\bf 98} (1985) 65.

\bibitem{ms} V. B. Matveev and M. A. Salle, {\em
     Darboux transformations and Solitons}, (Springer-Verlag, 1991).

\bibitem{mp} C. Morosi and L. Pizzocchero, {\em Commun. Math.
     Phys. } {\bf 176} (1996) 353.


\bibitem{o} W. Oevel, {\em Physica} {\bf A 195} (1993) 533;
 W. Oevel and W. Schief, Darboux transformations
    and KP Hierarchy, in {\em Applications of Analytic and Geometric
    Methods to Nonlinear Differential Equations}, ed. P. A.
    Clarkson, pp 193 (Kluwer Academic Publishers, 1993).

\bibitem{rk} G. H. M. Roelofs and P. H. M. Kersten, {\em J. Math.
Phys.} {\bf 33} (1992) 2185.

\bibitem{uy} K. Ueno and H. Yamada, Supersymmetric Extensions
    of the Kadomtsev-Petviashvili Hierarchy and the Universal 
    Super Grassmann Manifold, in {\em Advanced Studies in 
    Pure Mathematics} {\bf 16} (1988) 373.
\bibitem{wks} M. Wadati, H. Sanuki and K. Konno, {\em Prog. Theor.
    Phys.} {\bf 53} (1975) 419.

\end{thebibliography}
\end{document}